\documentclass[12pt]{iopart}

\usepackage{epsf}
\usepackage{graphicx}  
\begin{document}

\title[Direct photons $\sim$basis for characterizing heavy ion collisions$\sim$]{Direct photons $\sim$basis for characterizing heavy ion collisions$\sim$}

\author{Takao Sakaguchi}

\address{Brookhaven National Laboratory, Physics Department, Upton, NY 11973, U.S.A.}
\ead{takao@bnl.gov}
\begin{abstract}
After years of experimental and theoretical efforts, direct photons become
a strong and reliable tool to establish the basic characteristics of a hot
and dense matter produced in heavy ion collisions. The recent direct photon
measurements are reviewed and a future prospect is given.
\end{abstract}


\section{Introduction}
\begin{figure}[htb]
\centering
\begin{minipage}{75mm}
\includegraphics*[width=6.0cm]{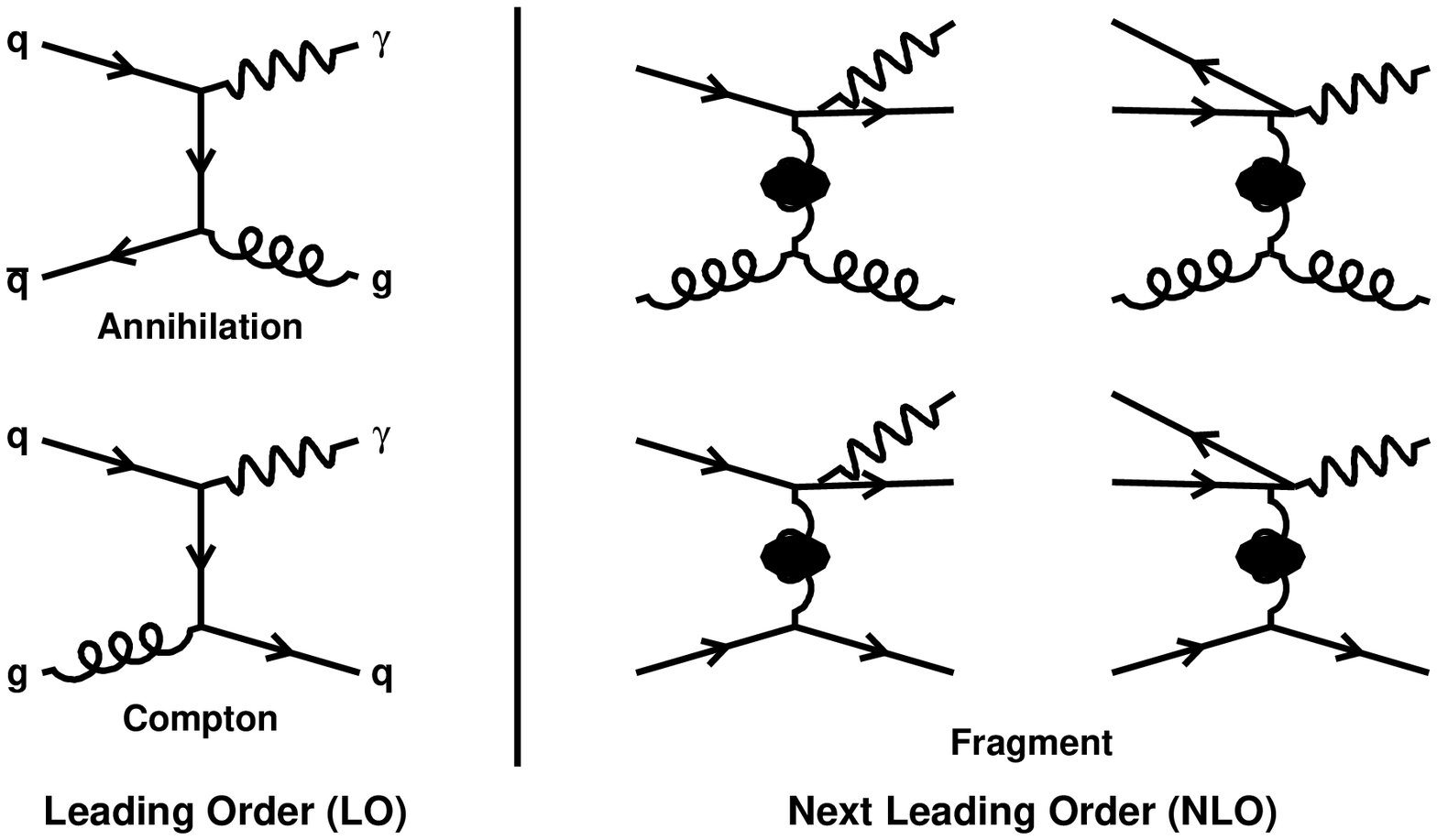}
\end{minipage}
\vspace{2mm}
\begin{minipage}{75mm}
\includegraphics*[width=6.0cm]{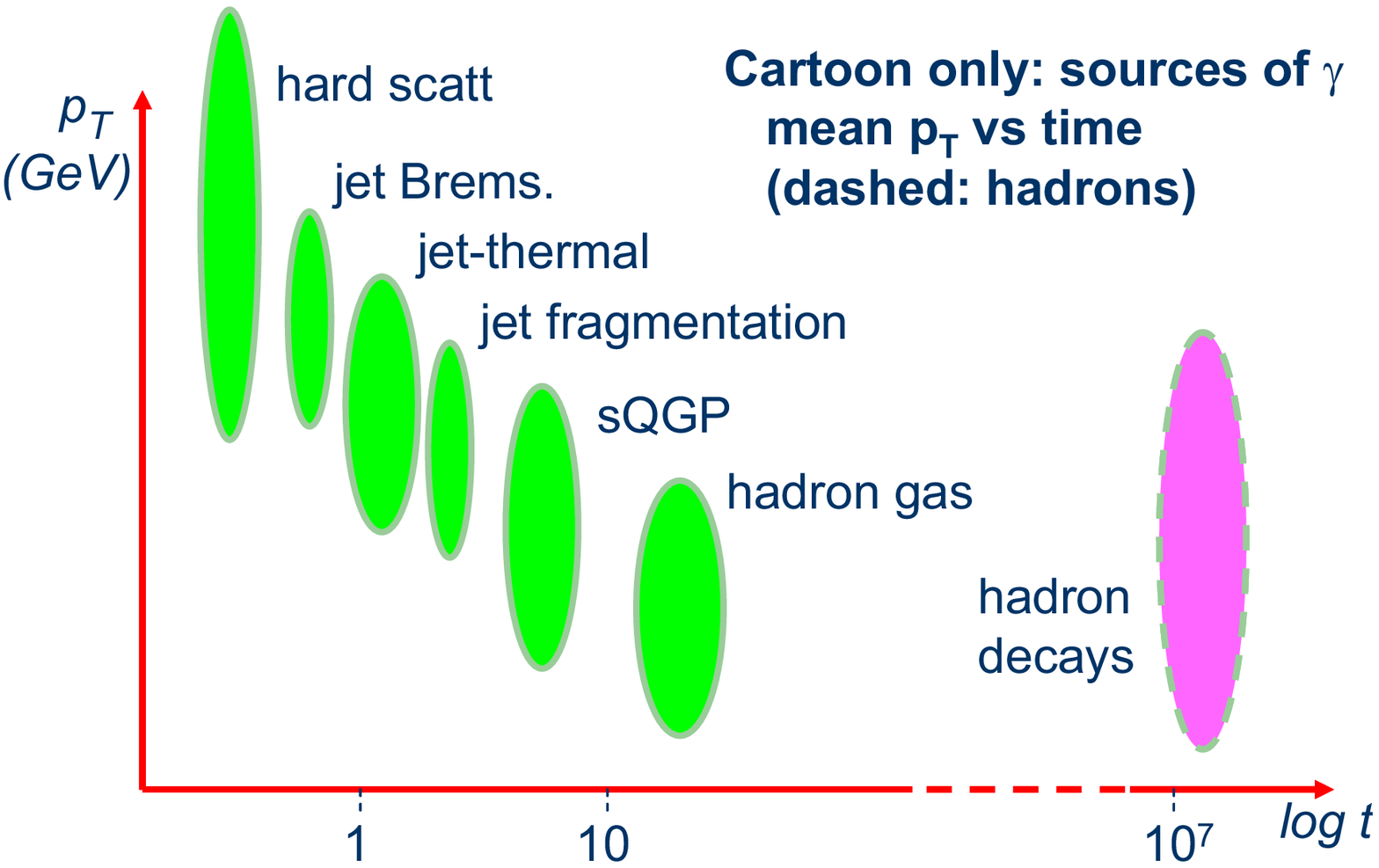}
\end{minipage}
\caption{(a)Production processes of direct photons (left), and (b) their manifestation as a function $p_T$ (right).}
\label{figBasic}
\end{figure}

Direct photons are an excellent probe for extracting thermodynamical
information of a matter produced in nucleus-nucleus collisions, as they are
emitted from all the stages of collisions, and don't interact strongly with
medium once produced. They are produced through a Compton scattering
of quarks and gluons ($qg\rightarrow q \gamma$) or an annihilation of
quarks and anti-quarks ($q\overline{q} \rightarrow g \gamma$) as leading
order processes, and the next leading order (NLO) process is dominated by
bremsstrahlung (fragment) ($qg \rightarrow qg\gamma$) as depicted in
Fig.~\ref{figBasic}(a). Theoretical studies show that a part of fragment
processes arises as leading order~\cite{ref32}. The source of radiations
are from various processes and manifest as a function of transverse
momentum ($p_T$)~\cite{ref1}(Fig.~\ref{figBasic}(b)).
Photons with high $p_T$ are primarily produced in the initial hard
scattering, and often called as "hard photons". Under the formation of hot
and dense medium, in addition to the hard photons, a calculation predicts
that the photon contribution from a quark gluon plasma (QGP) state
dominates lower transverse momentum ($p_T$) region in heavy ion
collisions (1$<p_T<$3\,GeV/$c$ in Au+Au collisions at
$\sqrt{s_{NN}}$=200\,GeV~\cite{ref2}). The signal from a hadron rescattering
process dominates even lower $p_T$. Compton scattering of hard-scattered
partons and partons in the medium (jet-photon conversion), or bremsstrahlung
of the hard scattered partons in the medium will also arise~\cite{ref3}.
Photons from the processes would become an another measure of the parton
density of the medium since they are produced through an interaction of
hard-scattered partons and the medium. All the contributions are overwhelmed
by huge photonic background from known hadron sources such as $\pi^{0}$'s
or $\eta$'s, except for high $p_T$.

\section{High $p_T$ direct photons $\sim$how well are they calibrated?$\sim$}
\subsection{$p+p$ collisions: precision test}
\begin{figure}[htb]
\centering
\begin{minipage}{65mm}
\includegraphics*[width=5.0cm]{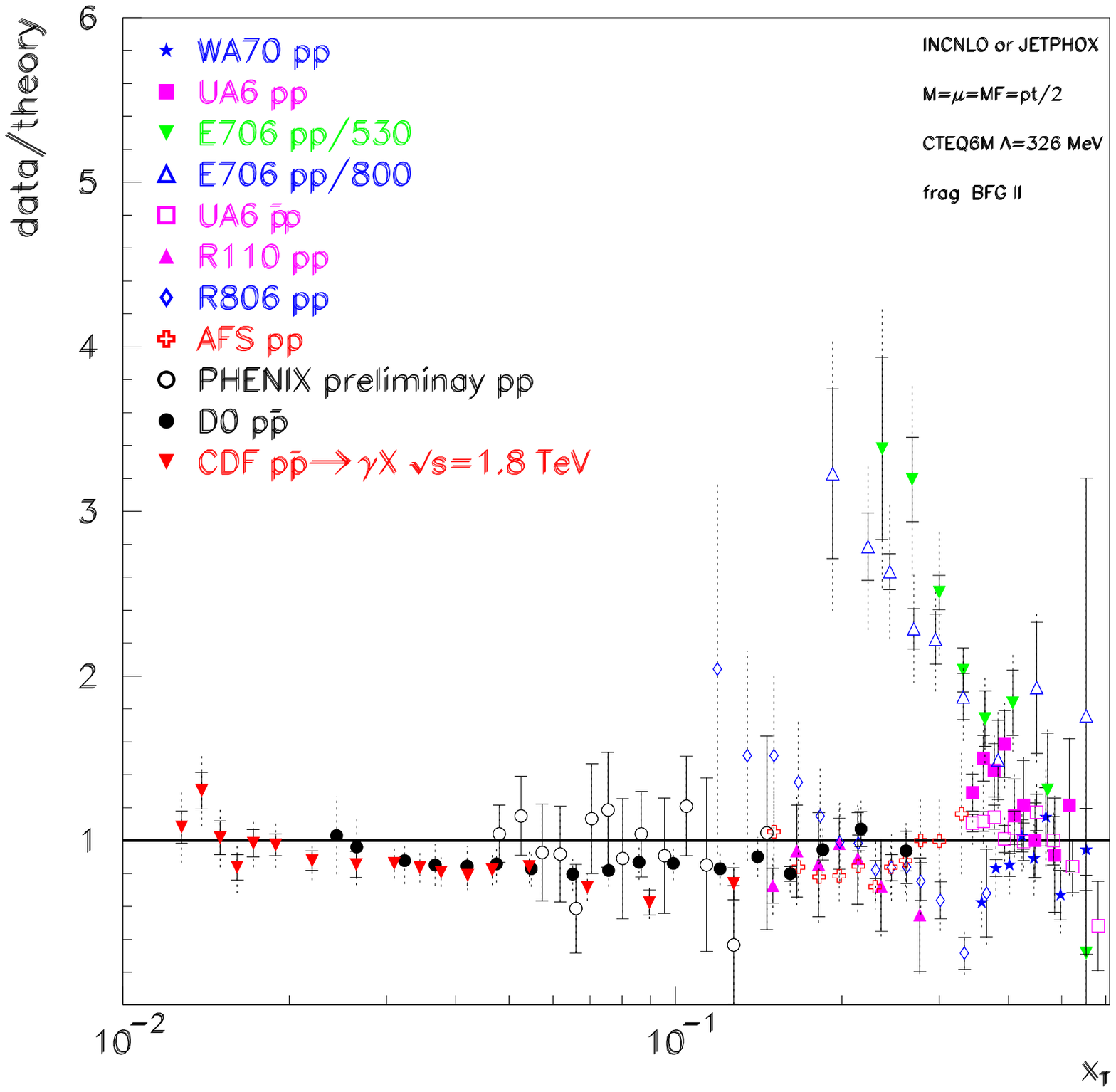}
\end{minipage}
\begin{minipage}{70mm}
\includegraphics*[width=5.5cm]{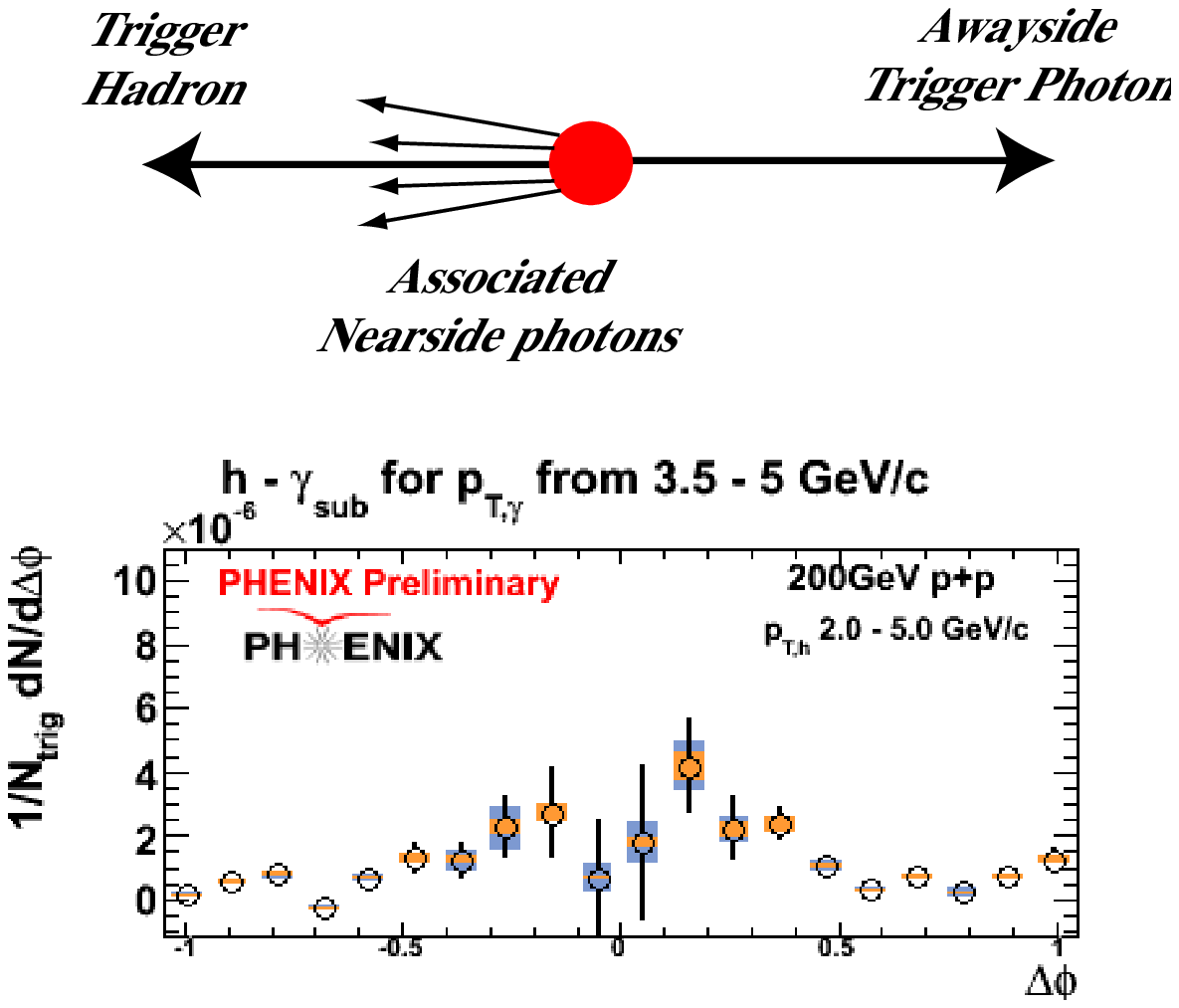}
\end{minipage}
\caption{(a)Data over NLO pQCD from various experiments in p+p collisions, and (b)fragment photons measured through $h-\gamma_{dir}$ $\Delta\phi$ correlation by PHENIX.}
\label{figAurenche}
\end{figure}

Hard direct photon production in p+p collisions is extensively studied both
experimentally and theoretically. Figure~\ref{figAurenche} shows the ratios
of direct photon cross-sections to NLO pQCD calculation measured by various
experiments~\cite{ref4}.
The data are explained by NLO pQCD calculations within $\sim$20\,\%.

As depicted in Fig.~\ref{figBasic}(a), photons are produced in the fragment
process as well. Several experiments have measured the prompt to all hard
photons by applying an isolation cut. The same cut is applied to a NLO pQCD
calculation to compare with the measurement. The PHENIX experiment at RHIC
has used a similar technique and confirmed that the calculation is consistent
with the result~\cite{ref6}. There is a new attempt of measuring fragment
contribution more directly. The PHENIX experiment has recently measured
photons associated with the same side of trigger high $p_T$
hadrons~\cite{ref7}. These photons are considered to be from fragment
processes. Fig.~\ref{figAurenche}(b) shows the associated photon yield
$\Delta\phi$ distributions from the analysis. The ratio of
near-side-associated fragment to inclusive photons is also measured, and
show a consistency with the previous PHENIX measurement~\cite{ref6}.
A detailed study of PID efficiency important in the analysis because
possible mis-identification of photons/hadrons would produce a trigger bias.

\subsection{Hard photons suppressed in Au+Au?}
\begin{figure}[htbp]
\centering
\begin{minipage}{5.0cm}
\includegraphics*[width=5.0cm]{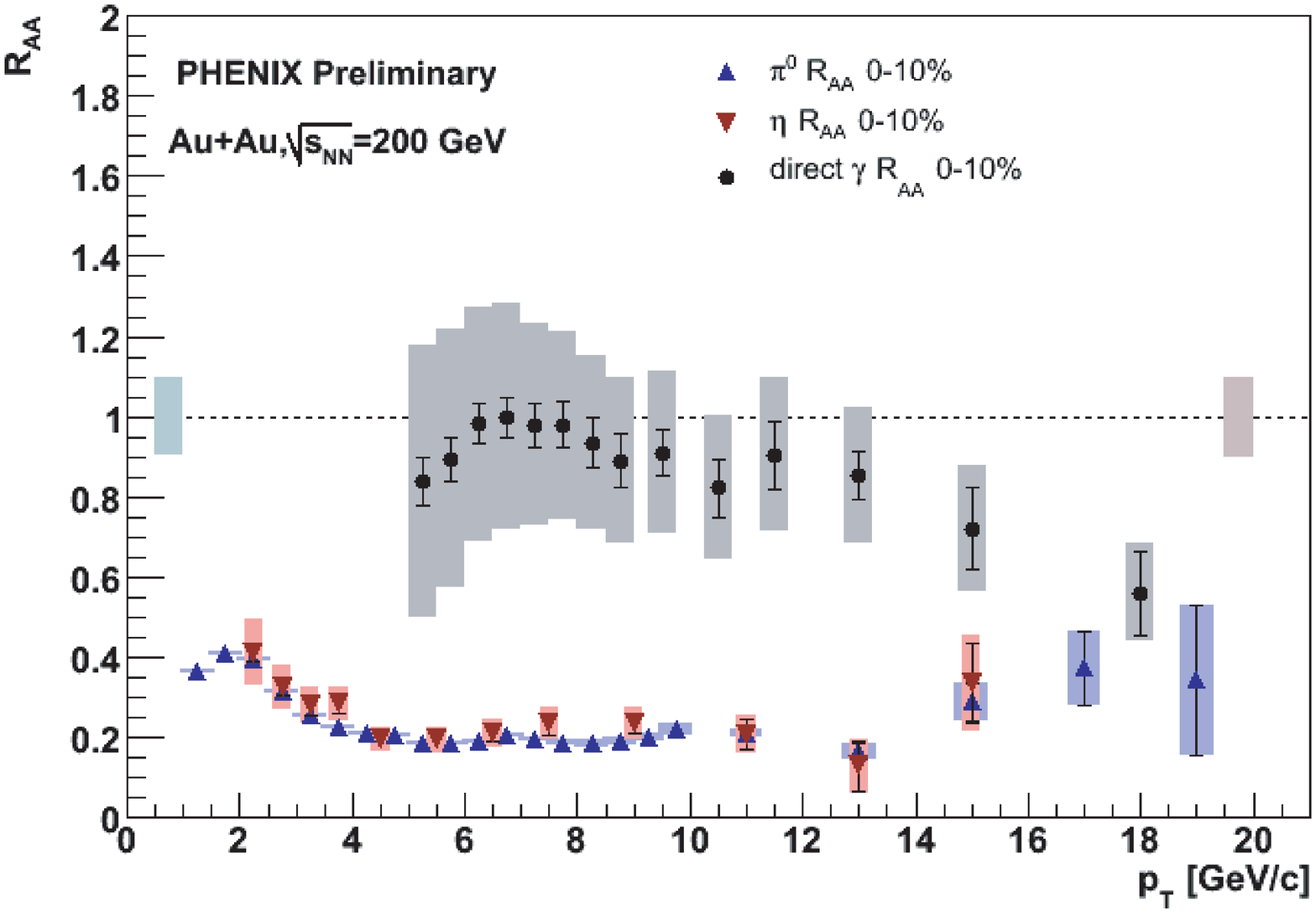}
\end{minipage}
\begin{minipage}{4.5cm}
\includegraphics*[width=4.5cm]{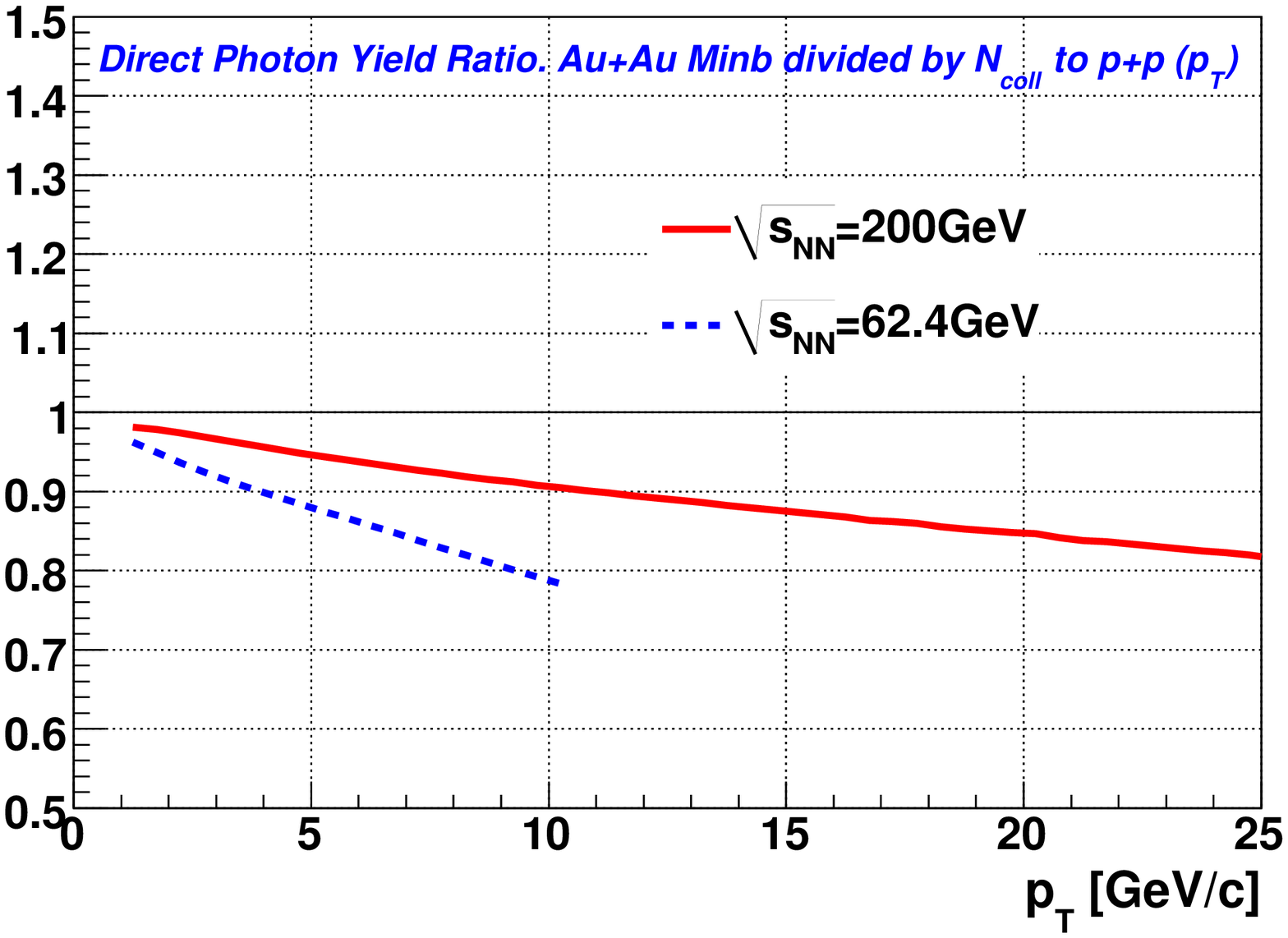}
\end{minipage}
\begin{minipage}{4.5cm}
\includegraphics*[width=4.5cm]{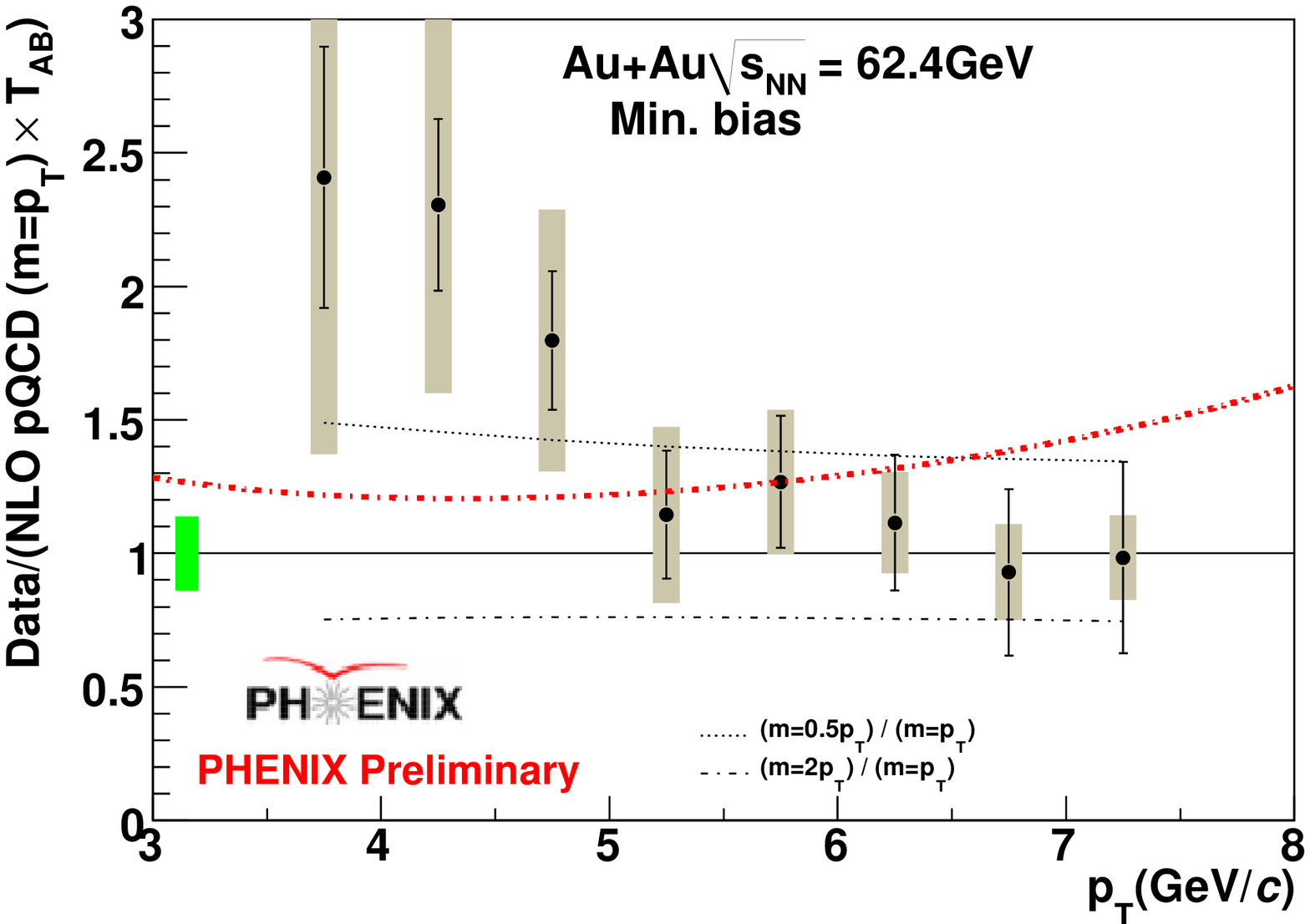}
\end{minipage}
\caption{(a) $R_{AA}$ for direct photons, $\pi^0$ and $\eta$ in Au+Au collisions at $\sqrt{s_{NN}}$=200GeV (left), (b) expected $R_{AA}$ from NLO pQCD calculation at 200 and 62\,GeV (middle), and (c) $R_{AA}$ for direct photons in Au+Au collisions at 62\,GeV (right) by PHENIX.}
\label{WernerCalc}
\end{figure}

The yield of high $p_T$ direct photons are well-scaled by a nuclear overlap
function ($T_{AB}$) in heavy ion collisions. The first measurement of such
photons at RHIC confirmed that the high $p_T$ hadron suppression is a
consequence of an energy loss of hard-scattered partons in the hot and dense
medium. The latest high statistics data from PHENIX showed a trend of
decreasing at high $p_T$ ($p_T>$14\,GeV/$c$) (Fig.~\ref{WernerCalc}(a)).
The decrease of the yield in Au+Au starts at $\sim$12\,GeV/$c$ ($x_T$=0.12)
and drops by $\sim$30\,\% at 18\,GeV/$c$ ($x_T$=0.18)~\cite{ref8}.
Parton distribution functions (PDFs) do not change by 30\,\% between
the two $x_T$ regions~\cite{ref9}.

Isospin effect has been proposed to explain the suppression~\cite{ref10}.
The photon production cross-section is proportional to
$\alpha \alpha_s \Sigma e_q^2$. Therefore, the yield of photons will be
different between p+p, p+n and n+n collisions. It results in the
deviation of $R_{AA}$ from unity at high $p_T$ in Au+Au collisions,
where the contribution of valence quarks become prominent. There is a
$\sim$15\,\% drop at 18\,GeV/$c$ at $\sqrt{s_{NN}}$=200\,GeV expected from
the effect (Fig.~\ref{WernerCalc}(b)). Combining PDF effect with the
isospin effect would explain the data. As shown in Fig.~\ref{WernerCalc}(b),
the effect will manifest in lower $p_T$ region at $\sqrt{s_{NN}}$=62.4\,GeV
because the effect scales with $x_T$. The PHENIX experiment has measured
photons in Au+Au collisions at $\sqrt{s_{NN}}$=62.4\,GeV~\cite{refSak}
and divided them by NLO pQCD instead of p+p yield (Fig.~\ref{WernerCalc}(c)),
since there is no p+p data from the experiment. The difference between p+p
yield and NLO pQCD calculation measured at 200\,GeV is scaled to 62.4\,GeV
and shown as a dot-dashed line in 62.4\,GeV. Assuming this is the baseline,
we may have confirmed the isospin effect (i.e., suppressed at
$p_T>$5\,GeV/$c$, corresponding to 16\,GeV/$c$ at 200\,GeV), also at 62.4\,GeV.
Combining the Au+Au data with the ones from future high statistics d+Au data
would disentangle the PDF and isospin effect.

\section{Application of well-calibrated probe $\sim$$\gamma$-jet analysis$\sim$}
\begin{figure}[htb]
\centering
\begin{minipage}{75mm}
\centering
\includegraphics*[width=5.5cm]{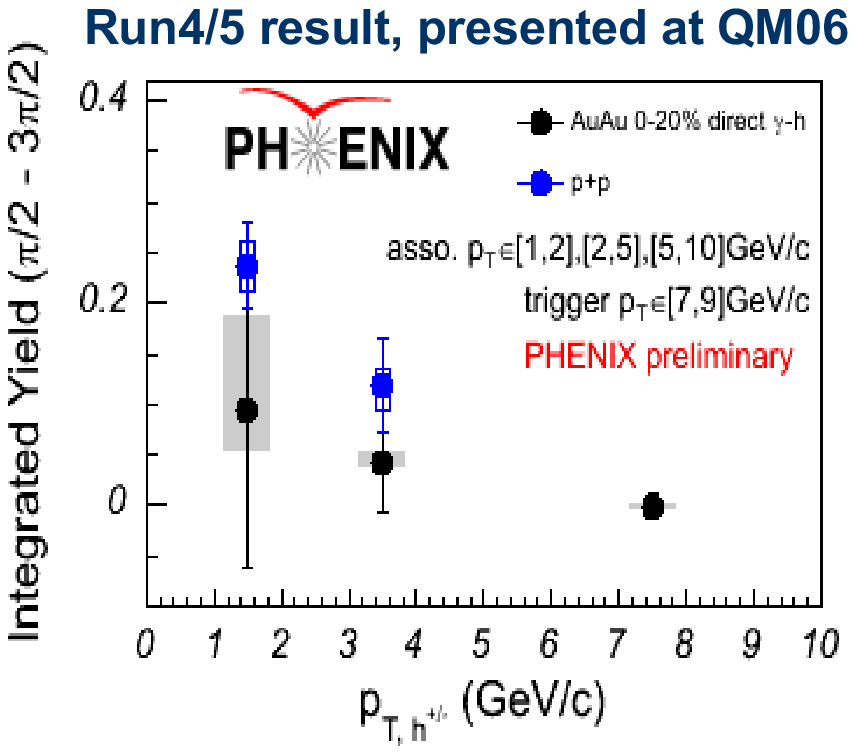}
\end{minipage}
\begin{minipage}{75mm}
\centering
\includegraphics*[width=5.5cm]{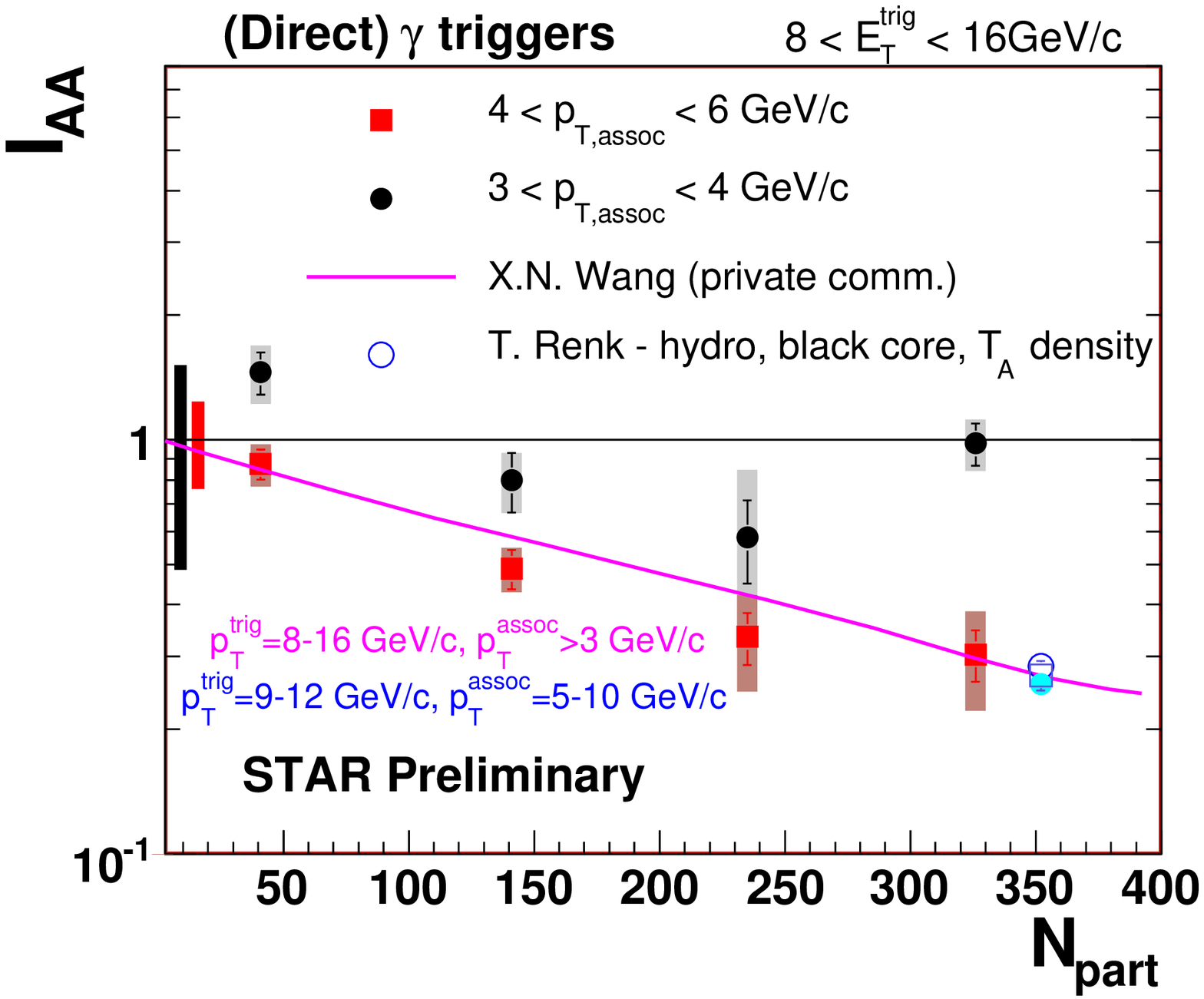}
\end{minipage}
\caption{Away-side associated hadron yield measured in $\gamma_{dir}-h$ correlation in p+p and Au+Au collisions at (a) PHENIX (left) and (b) STAR (right).}
\label{figFragMatt}
\end{figure}

Well calibrated high $p_T$ photons are ideal as measure of the initial
momenta of back-scattered partons. The idea was first proposed a decade
ago~\cite{ref29}, but the measurement has not become realized until recent.
The PHENIX has measured an associated away-side hadron yield when triggered
by a hard photon, both p+p and Au+Au collisions at $\sqrt{s_{NN}}$=200\,GeV
as shown in Fig.~\ref{figFragMatt}(a).
The result shows that the away-side per-trigger hadron yield is reduced in
Au+Au collisions compared to that expected from p+p collisions~\cite{ref7}.
This is qualitatively consistent with single particle measurement~\cite{ref8}.
In this analysis, all the hadron contribution associated with photons
from hadron decay are subtracted on statistical basis to obtain
$\gamma_{dir}-h$ correlation as:
\[ (\gamma_{dir}-h) = (\gamma_{incl.}-h) - (\gamma_{dec}-h) \]
There is a new result from the STAR experiment showing $I_{AA}$ in
$\gamma_{dir}-h$ correlation~\cite{ref14}. In this analysis, the correlation
is obtained by:
\[ (\gamma_{dir}-h) = (Clus_{\gamma-en}-h) - \alpha\times(\pi^{0}-h) \]
where $Clus_{\gamma-en}$ stands for $\gamma$-enriched clusters by
a shower shape cut. $\alpha$ is determined such that the near side associated
hadron yield be zero. This procedure will be justified under the assumption
that the $\eta$-triggered hadron yield and fragment-photon-triggered hadron
yield are as same as the $\pi^0$-triggered hadron yield. The cross-section
measurement of direct photons would help justifying the procedure.

\section{Thermal photons from CERN to RHIC}
The measurement of thermal photons delivers the temperature of the system.
Combining the temperature with an entropy derived from a particle multiplicity
measurement will deduce the degree of the freedom of the system~\cite{ref30,ref33}.
There is a direct photon measurement in thermal region in Pb+Pb collisions,
made by WA98 experiments at CERN~\cite{ref17}. However, the lack of p+p
measurement at the same energy made it difficult to understand whether
or not there is a thermal emission~\cite{ref2}. WA98 has recently
analyzed p+Pb and p+C collision data to measure hard photons with
a nuclear effect ($k_T$ smearing). Taking the ratio of the yield in
Pb+Pb to p+Pb should be able to quantify the pure non-hard photon component.
However, the error is too large to make a conclusion. The experiment
is now making an effort to minimize the systematic errors~\cite{ref18}.
\begin{figure}[htb]
\centering
\hspace{-5mm}
\begin{minipage}{6.0cm}
\includegraphics*[width=5.5cm]{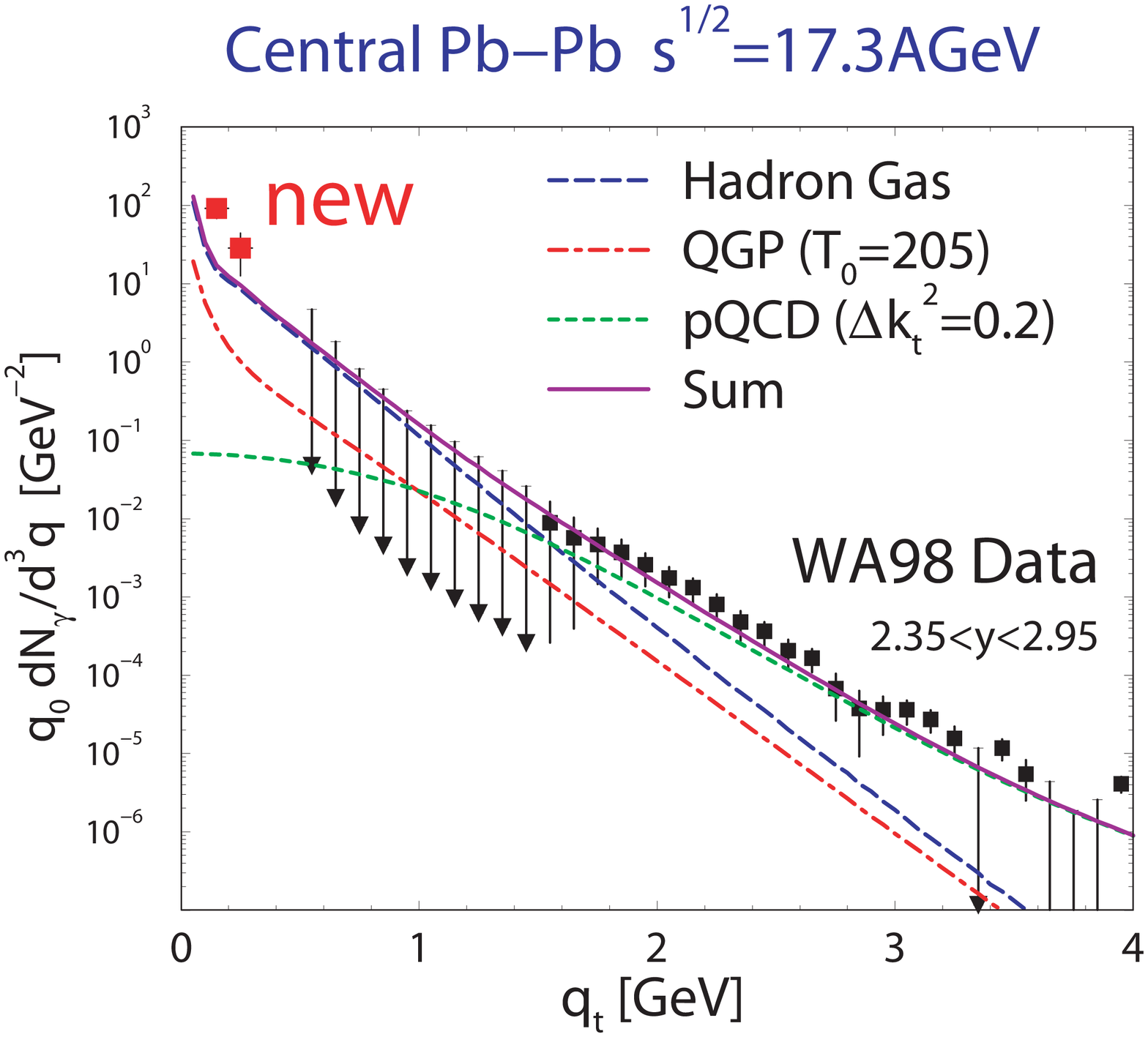}
\end{minipage}
\hspace{8mm}
\begin{minipage}{8.2cm}
\includegraphics*[width=7.4cm]{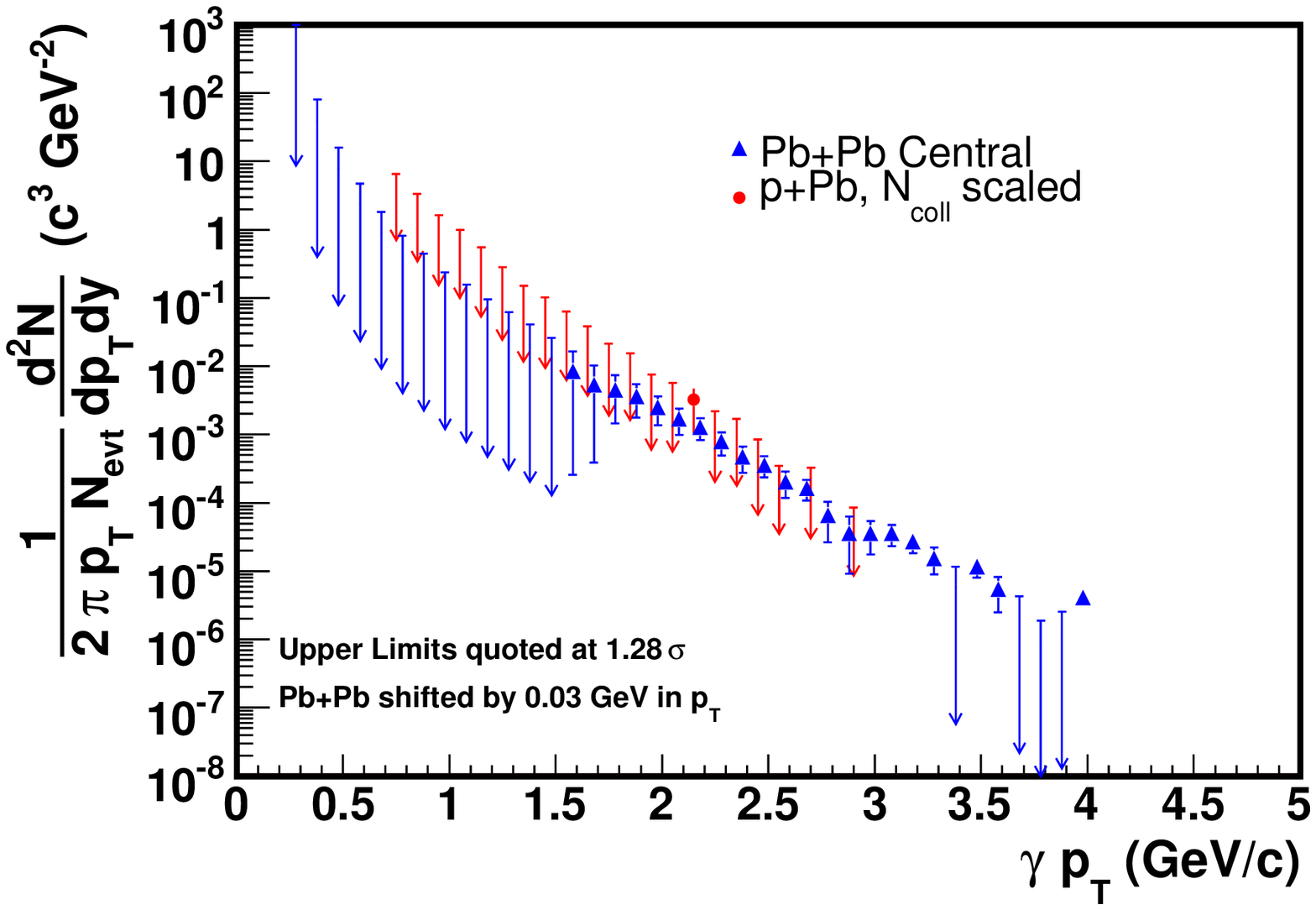}
\end{minipage}
\caption{(a) Direct photon measurement in thermal region by WA98 at CERN and theoretical interpretation (left), and (b) comparison of direct photon yield in p+Pb and Pb+Pb (right).}
\end{figure}

\begin{figure}[htbp]
\centering
\begin{minipage}{7cm}
\includegraphics*[width=4.5cm]{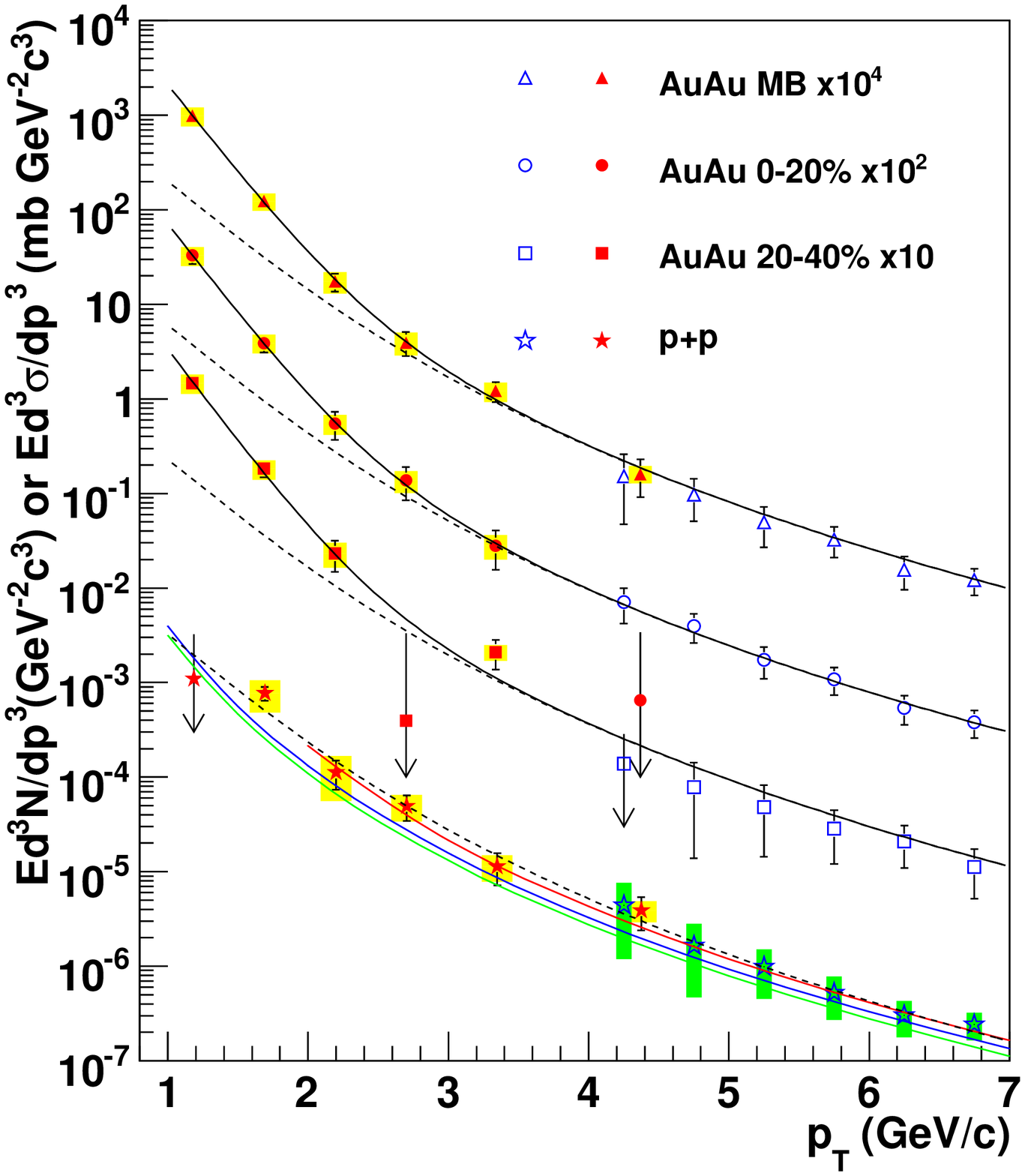}
\end{minipage}
\begin{minipage}{7cm}
\includegraphics*[width=7.0cm]{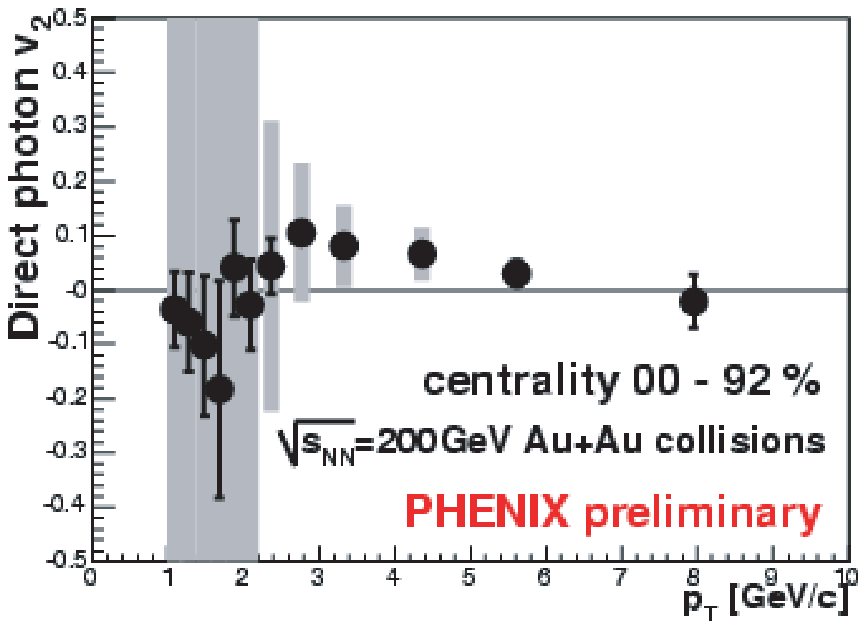}
\end{minipage}
\caption{(a) Low $p_T$ direct photon spectra (left) and (b) direct photon elliptic flow (right) in Au+Au collisions measured by the PHENIX experiment at RHIC}
\label{figIntv2}
\end{figure}

The thermal photon contribution is believed to be $\sim$10\,\% at RHIC energy,
and might need a measurement with an error of $<$5\,\%. Measurement of the
internal conversion of direct photons
($\gamma \rightarrow \gamma^{*} \rightarrow e^{+}e^-$) opened up a possibility
to significantly reduce the systematic errors. The PHENIX experiment has
applied the technique of measuring low $p_T$ and low mass di-electrons to
high $p_T$ and low mass di-electrons. The measured yield is converted into
a direct photon yield using Kroll-Wada formula~\cite{ref19,ref20}.
If $M_{ee}\ll p_{T}$, and $M_{ee}<2M_{\pi}$, there is little contribution
from $q\overline{q} \rightarrow \gamma^{*}$, and thus the conversion is
straightforward. The yield of direct photons are found to be higher than
the ones expected from p+p collisions scaled by the number of binary
collisions, suggesting there are additional sources of photons in Au+Au
system (Fig.~\ref{figIntv2}(a)).
The average of simple exponential fits to the low $p_T$ regions gives
a temperature of 220$\pm$23$\pm$8\,MeV. However, the possible contribution
from a nuclear effect ($k_T$ smearing) to the $p_T$ region still
remains~\cite{ref28}. The internal conversion technique would help precisely
determining the contributions in d+Au collisions. It should be noted that
there are a number of theoretical analyses for extracting thermodynamical
quantities~\cite{ref33,ref34}.

\section{Decomposition of photon sources --photon elliptic flow--}
It is predicted that the elliptic flow ($v_2$) of photons show the different
sign and/or magnitude, depending on the production processes of
photons~\cite{ref21}. The observable is powerful to disentangle the
contributions from various photon sources in the $p_T$ region where they
intermixes. The photons from hadron-gas interaction and thermal radiation
follow the collective expansion of a system, and would give a positive $v_2$.
The amount of photons produced by jet-photon conversion or in-medium
bremsstrahlung increases as the medium to traverse increases.
Therefore these photons show a negative $v_2$. The intrinsic fragment or
bremsstrahlung photons will give positive $v_{2}$ since larger energy
loss of jets is expected in out-plane.

PHENIX has measured the $v_2$ of direct photons by subtracting the $v_2$
of hadron decay photons off from that of the inclusive photons, following
the formula below:
\[ {v_2}^{dir.} = (R \times {v_2}^{incl.} -{v_2}^{bkgd.})/(R-1),\ \ \ R = (\gamma/\pi^0)_{meas}/(\gamma/\pi^0)_{bkgd}\]
The result is shown in Fig.~\ref{figIntv2}(b).
Although the systematic error is very large, the $v_2$ of direct photons
tends to be positive in 3-6\,GeV/$c$ independent of centrality~\cite{ref22}.
It naively implies that the contribution from intrinsic fragment or
bremsstrahlung photons are dominant over that from jet-photon conversion
process. It could happen if the energy loss is very large and most of the
hadrons observed are produced near surface of the system; hard scattered
partons are absorbed before making enough Compton scattering to produce
additional photons. In any case, minimizing the systematic error is desired
before making a conclusion.

\section{What would be the next measurement?}
\subsection{LHC}
At LHC energies, the cross-section of hard photons increases drastically,
and therefore the primary target will be to measure the energy loss of
hard scattered partons with a trigger of prompt photons; the measurement
of $\gamma$-jet correlation~\cite{ref23}.
\begin{figure}[htbp]
\centering
\begin{minipage}{5.5cm}
\includegraphics*[width=5.5cm]{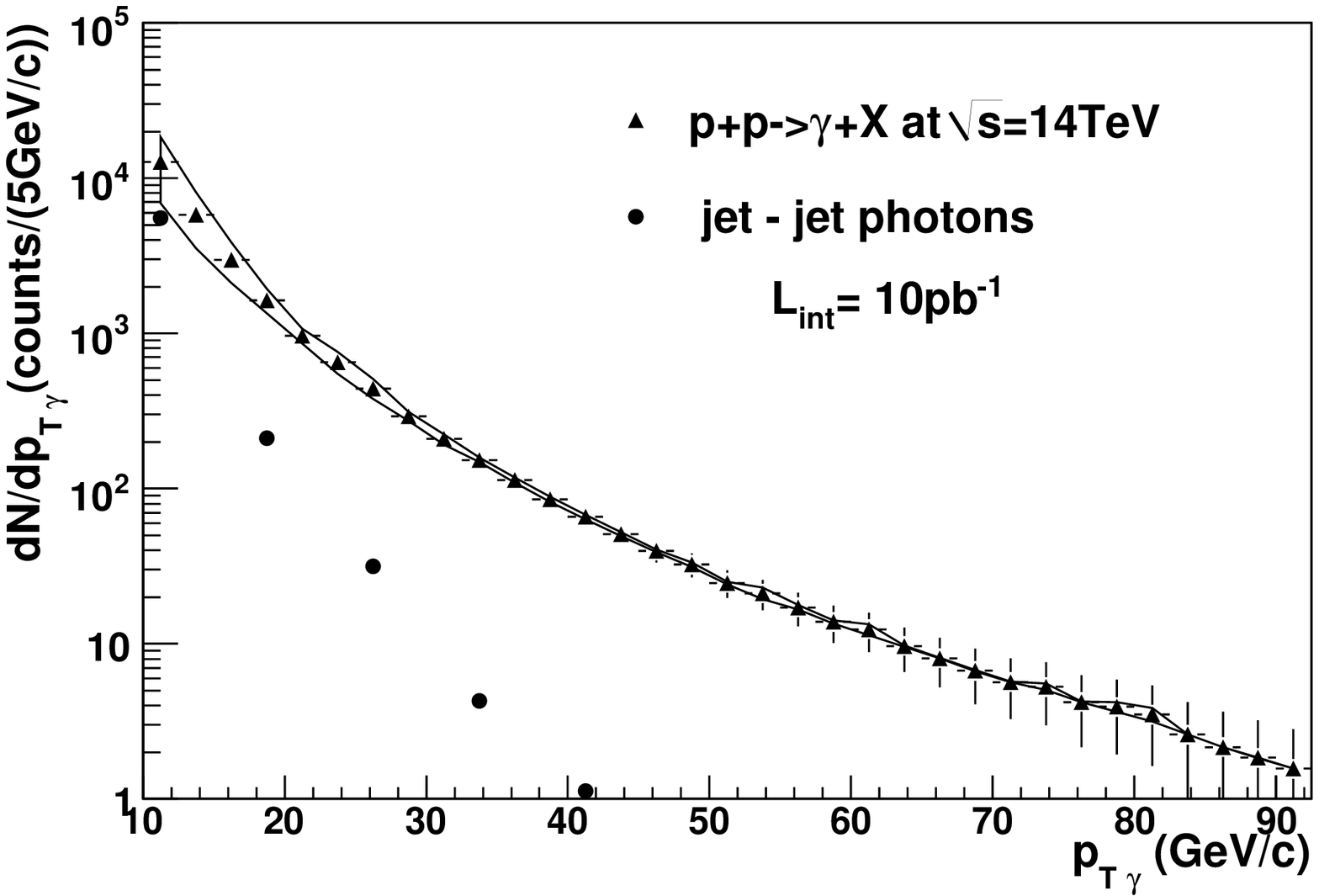}
\end{minipage}
\hspace{-5mm}
\begin{minipage}{4.0cm}
\includegraphics*[width=4.0cm]{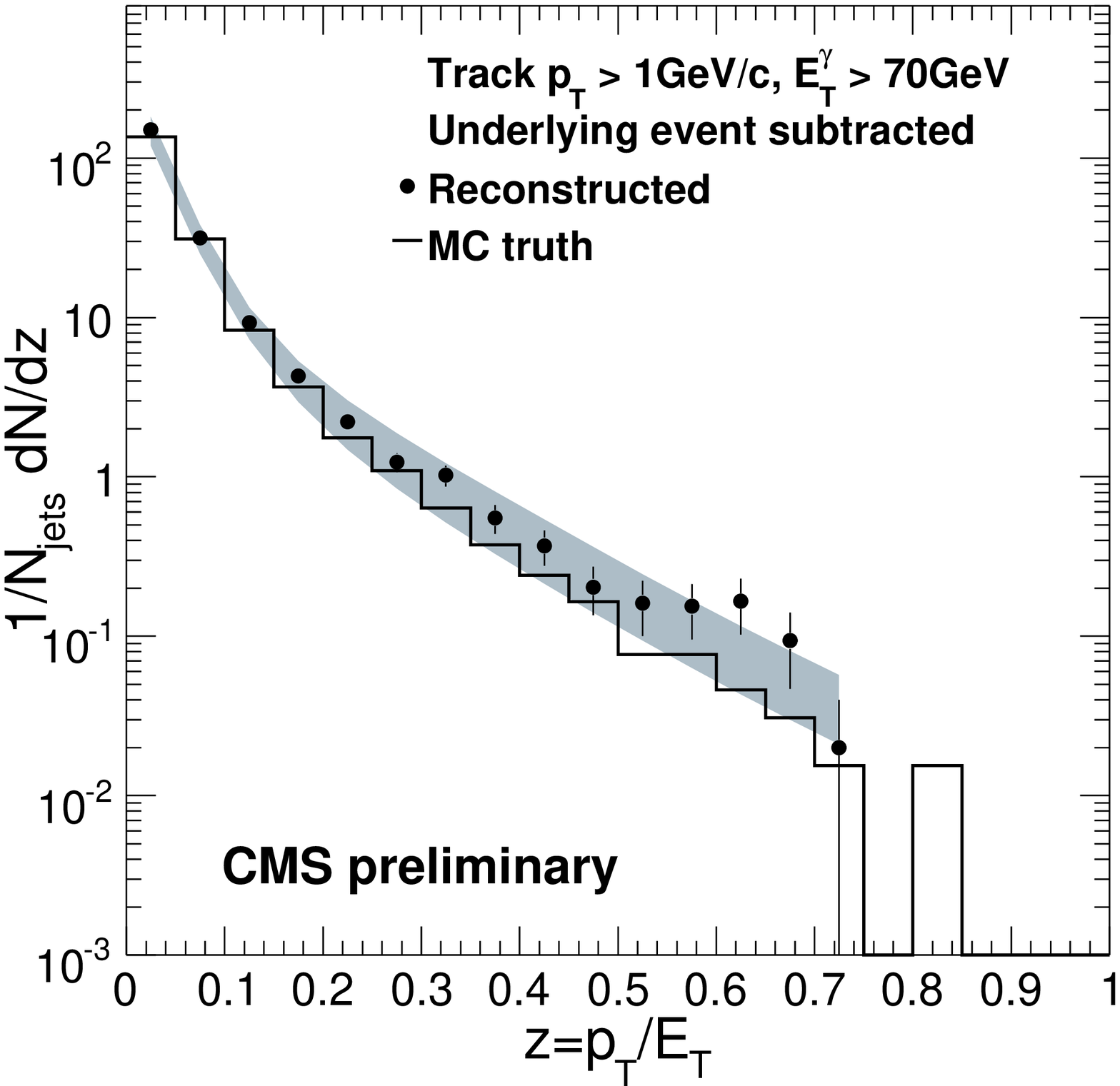}
\end{minipage}
\hspace{5mm}
\begin{minipage}{5.0cm}
\includegraphics*[width=4.5cm]{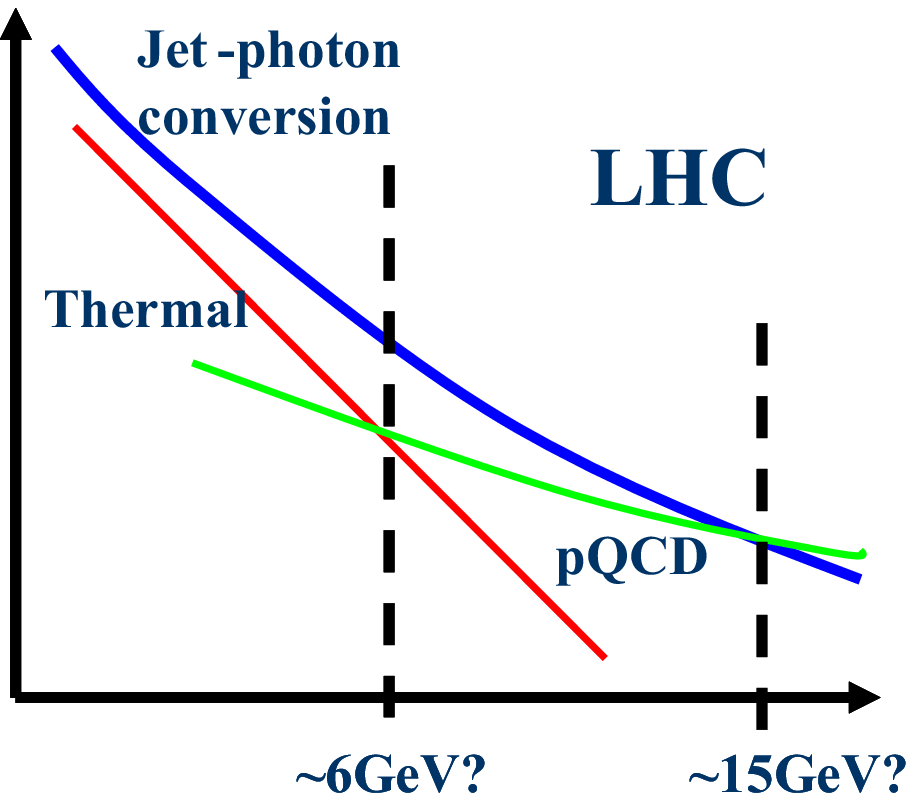}
\end{minipage}
\caption{(a) ALICE simulation of tagging efficiency of an optimized cut for photons (triangle), and rejection power to hadrons (circle) (left), (b) CMS simulation of fragmentation function reconstruction (middle), and (c) possible scenario of photon contributions at low to mid $p_T$ region at LHC (right).}
\label{figALICECMS}
\end{figure}
The experiments planning the measurement at LHC have already started
feasibility studies on the measurement using realistic simulations.
Figure~\ref{figALICECMS}(a) shows the tagging efficiency of prompt
photons and rejection power to hadrons with an optimized cut in the ALICE
detector~\cite{ref24}. At high $p_T$, the tagged samples
are shown to be mostly photons. Figure~\ref{figALICECMS}(b) shows the
reconstructed fragmentation function in CMS detector~\cite{ref25}.
The function is in good agreement with the
input fragmentation function within systematic errors in hand.
These studies show that measurement of energy loss of partons in the medium
is promising at LHC.

Turing eyes into the low to mid $p_T$ region, thermal photon emission would
be of great interest as are at CERN and RHIC. Here, we can estimate how well
such photons are resolved. The temperature and the degree of freedom can be
associated with energy density. Using the fact and following the idea
presented in ~\cite{ref2}, the yield of thermal photons can simply be
written as:
\[\sigma\sim N_{part}\times(\tau_{freeze}-\tau_{0})\times(s^{1/2})^{1/4\times2}\]
At LHC, the c.m.s. energy will increase by a factor of 70 and the $N_{part}$
by a little, therefore the yield may increase by a factor of $\sim$9.
On the other hand, photons related to hard-scattered partons would
increase drastically in LHC, because the jet cross-section becomes
exponentially larger as a function of c.m.s. energy.
The jet-photon conversion yield would be proportional to the multiple of
jet cross-section and QGP volume, resulting in:
\[\sigma \sim N_{part} \times N_{coll} \times (s^{1/2})^n \times g(x_T)\]
where the last term represents hard-scattering cross-section, $n$ is the
$x_T$-scaling power, and $\sim5-8$. Therefore, the jet-photon conversion
would overwhelm thermal photon production. A rough schematics is shown
in Fig.~\ref{figALICECMS}(c). From these consideration, it would be hard to
observe thermal photons, instead, the medium can be investigated by observing
photons from the jet-photon conversion process, together with $v_2$ measurement.

\subsection{Forward measurement}
\begin{figure}[htbp]
\centering
\begin{minipage}{7cm}
\includegraphics*[width=5.0cm]{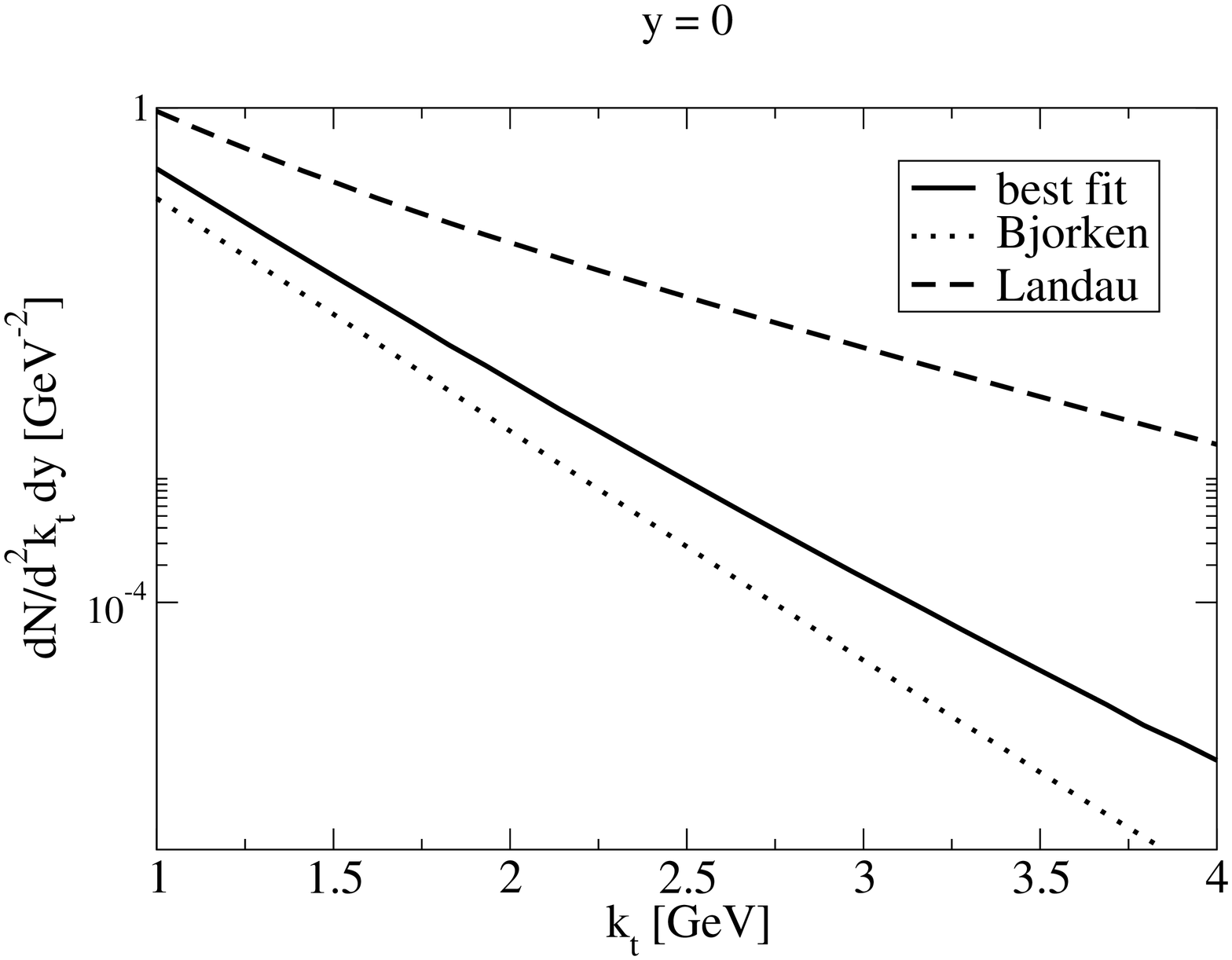}
\end{minipage}
\begin{minipage}{7cm}
\includegraphics*[width=5.0cm]{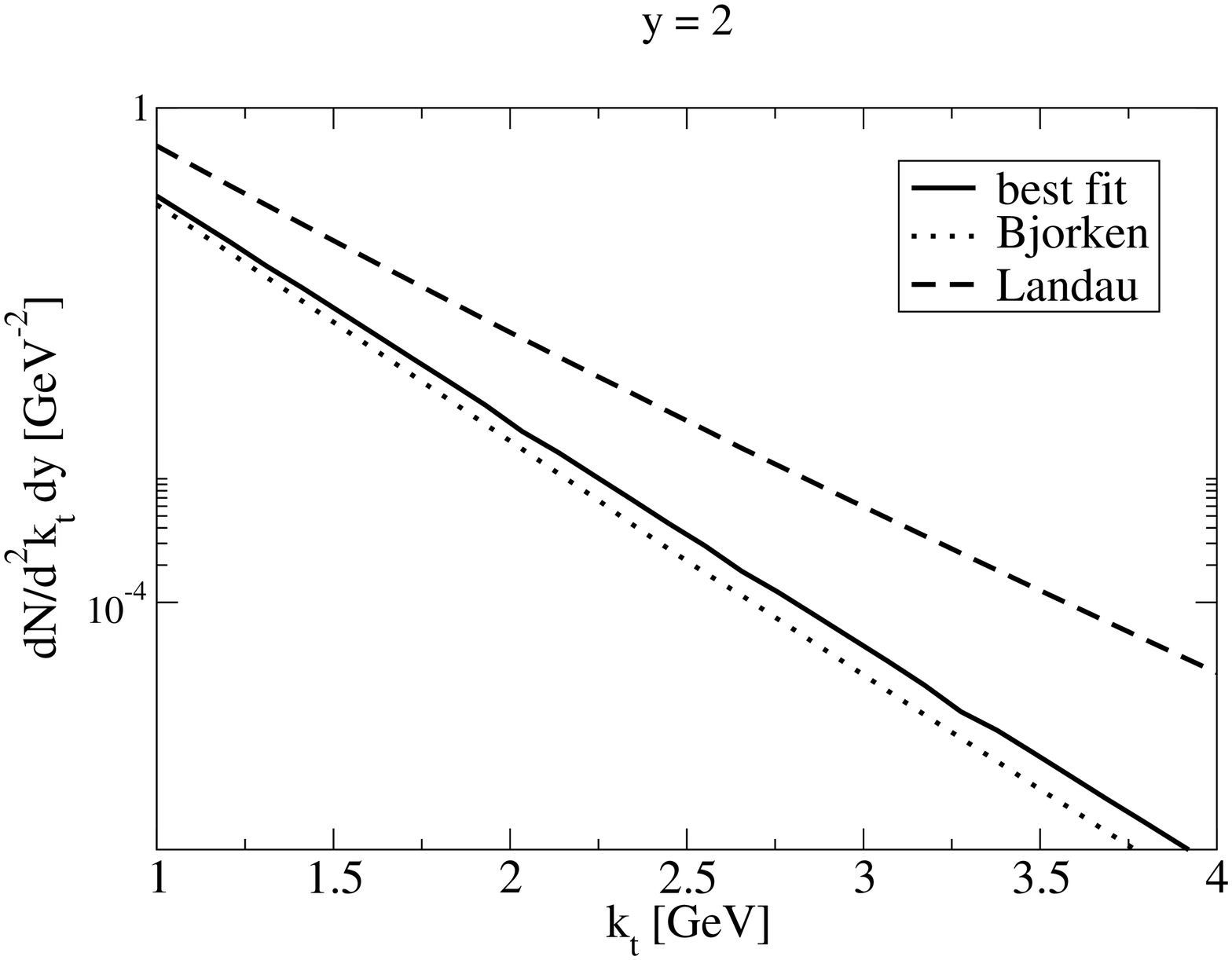}
\end{minipage}
\caption{Direct photon cross-section at (a)y=0 (left) and (b)y=2 (right), under different system expansion scenario.}
\label{figRenk}
\end{figure}

Comparison of the hadron production at mid and forward rapidities has
deduced particle production mechanisms, such as CGC. There is an
interesting prediction on photon production at mid and forward rapidity,
which can discriminate system
expansion scenarios as shown in Fig.~\ref{figRenk}~\cite{ref26}.
Landau and Bj\"{o}rken expansion of the system
would differ the ratio of the yield in mid and forward rapidity. At RHIC,
STAR has photon detector~\cite{ref27} and has already measured photons at a
forward rapidity. PHENIX also has a photon detector at the
rapidity~\cite{ref31}. The detectors would provide interesting results
on the system expansion.

\section{Summary}
The recent direct photon measurements were reviewed and a future prospect
was given. Direct photons would establish a status as one of the most
fundamental measurement in heavy ion collisions in the future.


\section*{References}
\begin{thebibliography}{99}
\bibitem{ref32} P.~Arnold, G.~D.~Moore and L.~G.~Yaffe, {\it JHEP} {\bf 0111}, 057 (2001).
\bibitem{ref1} G. David, {\it Nucl. Phys.} {\bf A783} (2007) 359.
\bibitem{ref2} S. Turbide, R. Rapp and C. Gale, {\it Phys. Rev.} {\bf C69}, 140903 (2004).
\bibitem{ref3} R. Fries et al., {\it Phys. Rev.} {\bf C72}, 041902 (2005).
\bibitem{ref4} P. Aurenche, et al., {\it Phys. Rev.} {\bf D73}, 094007 (2006).
\bibitem{ref6} S. S. Adler et al. (PHENIX Coll.), {\it Phys. Rev. Lett.} {\bf 98}, 012002 (2007).
\bibitem{ref7} M. Nguyen (PHENIX Coll.), these proceedings.
\bibitem{ref8} T. Isobe (PHENIX Coll.), {\it J.\ Phys.\ G} {\bf 34} (2007) S1015.
\bibitem{ref9} K.J. Eskola, V.J. Kolhinen and C.A. Salgado, {\it Eur. Phys. J.} {\bf C9}, 61 (1999).
\bibitem{ref10} F. Arleo, {\it JHEP} {\bf 0609}(2006)015.
\bibitem{refSak} T. Sakaguchi (PHENIX Coll.), {\it Nucl. Phys.} {\bf A805} (2008) 355c.
\bibitem{ref29} X.~N.~Wang and Z.~Huang, {\it Phys. Rev.} {\bf C55}, 3047 (1997).
\bibitem{ref14} A. Hamed et al. (STAR Coll.), these proceedings.
\bibitem{ref30} B.~Muller and K.~Rajagopal, {\it Eur.\ Phys.\ J.} {\bf C43}, 15 (2005).
\bibitem{ref33} D.~d'Enterria and D.~Peressounko,{\it Eur.\ Phys.\ J.} {\bf C46}, 451 (2006).
\bibitem{ref17} M. M. Aggarwal et al. (WA98 Coll.), {\it Phys. Rev. Lett.} {\bf 85} (2000) 3595.
\bibitem{ref18} C. Baumann (WA98 Coll.), these proceedings.
\bibitem{ref19} T. Dahms (PHENIX Coll.), these proceedings; A.~Adare et al., arXiv:0804.4168 [nucl-ex].
\bibitem{ref20} N.~M.~Kroll and W.~Wada, {\it Phys. Rev.} {\bf 98}, 1355 (1955).
\bibitem{ref28} M.~J.~Russcher (STAR Coll.), {\it J.\ Phys.\ G} {\bf 34}, S1033 (2007); D.~Peressounko (PHENIX Coll.), {\it J.\ Phys.\ G} {\bf 34}, S869 (2007).
\bibitem{ref34} J.~e.~Alam, et al., {\it J.\ Phys.\ G} {\bf 34}, 871 (2007).
\bibitem{ref21} S. Turbide, C, Gale and R.J. Fries, {\it Phys. Rev. Lett.} {\bf 96}, 032303 (2006); R. Chatterjee et al., {\it Phys. Rev. Lett.} {\bf 96}, 202302 (2006); S. Turbide et al., arXiv:0712.0732
\bibitem{ref22} K. Miki (PHENIX Coll.), these proceedings.
\bibitem{ref23} S. Abreu et al., {\it J. Phys. G: Nucl. Part. Phys.} {\bf 35} (2008) 054001.
\bibitem{ref24} A. Morsch (ALICE Coll.), these proceedings. 
\bibitem{ref25} C. Loizides (CMS Coll.), these proceedings; arXiv:0804.3679[nucl-ex].
\bibitem{ref26} T. Renk, {\it Phys. Rev.} {\bf C71}, 064905 (2005).
\bibitem{ref27} R. Raniwara (STAR Coll.), these proceedings.
\bibitem{ref31} E. Kistenev (PHENIX Coll.), these proceedings.
\end{thebibliography}
\end{document}